# Adversarial Robustness of Phishing Email Detection: A Comparative Study of TF-IDF + Logistic Regression and Fine-Tuned DistilBERT

**Tanveer Ahmed**[1] and **Seyedali Pourmoafi**[1]

[1]*School of Physics, Engineering and Computer Science, University of Hertfordshire, Hatfield, Herts AL10 9AB, United Kingdom*

Corresponding authors: tanveer.ahmed1188@gmail.com · S.Pourmoafi@herts.ac.uk

**Abstract**

Phishing emails remain one of the most persistent cybersecurity threats, and machine-learning classifiers are widely used to detect them. Most reported detection accuracies, however, are measured on clean, in-distribution test data rather than on emails deliberately altered to evade detection. This paper reports a controlled, pairwise comparison of two phishing-detection approaches a TF-IDF + Logistic Regression baseline and a fine-tuned DistilBERT transformer trained on a unified corpus of 82,255 emails drawn from six public datasets and evaluated under three conditions: normal in-distribution, synthetic phishing, and adversarial phishing. Both models exceeded 98% accuracy on clean data yet degraded sharply under adversarial testing: TF-IDF + LR fell to 64.00% (a 34.59-percentage-point drop) and DistilBERT fell to 63.64% (a 35.40-percentage-point drop) a gap of only 0.36 percentage points, equivalent to a single email in the 275-sample adversarial test set. LIME, SHAP, and attention-rollout analysis indicate the two models relied on different evidence yet showed similar vulnerability. Pairwise error analysis shows the models agreed on 54.9% of adversarial samples but each made a similar number of exclusive errors (24 and 25 respectively), indicating partly complementary rather than identical failure modes. The results show that clean-data accuracy does not predict adversarial robustness, and that adversarial testing should be a standard part of phishing-detection evaluation.

## 1. Introduction

Phishing remains one of the most common and damaging forms of cyber-attack because it targets human trust as well as technical systems: the Anti-Phishing Working Group recorded 1,270,883 phishing attacks in the third quarter of 2022 alone, at the time its highest quarterly total on record (APWG, 2022). Chiew, Yong and Tan (2018) describe this as a continuous “cat-and-mouse” contest between attackers and defenders, in which each new detection method is met with a new evasion technique. The problem has become harder as large language models make attacker-generated text more convincing: Bethany et al. (2025) found LLM-generated phishing can perform comparably to professionally written phishing in a large organisational setting, while Eze and Shamir (2024) report that AI-generated phishing emails are longer and more grammatically regular than earlier human-written examples, which may reduce the value of simple surface-level detection cues.

Meanwhile, published detection results continue to look very strong under clean test conditions for example, Somesha et al. (2022) report 99.39% accuracy for a TF-IDF-based classifier and Atawneh and Aljehani (2023) report 99.61% for a BERT-based model. These figures describe performance on held-out data drawn from the same distribution as training data; they say little about what happens once an attacker deliberately rewords a message to evade detection, even though the adversarial machine learning literature has repeatedly shown that small, targeted input changes can significantly degrade NLP model performance (Goodfellow, Shlens and Szegedy, 2015; Alzantot et al., 2018; Jin et al., 2020). Despite this, there has been comparatively little controlled, head-to-head evaluation of how a lexical, frequency-based classifier and a contextual transformer classifier behave under identical adversarial phishing conditions.

This paper addresses that gap with a controlled comparison of a TF-IDF + Logistic Regression model and a fine-tuned DistilBERT model, trained on the same data and evaluated on the same normal, synthetic, and adversarial test sets. The guiding research question is: *how does the adversarial robustness of a conventional TF-IDF + Logistic Regression model compare with that of a fine-tuned DistilBERT transformer when both are evaluated on the same manipulated phishing emails?* The contribution is not a new model architecture but an evaluation design: a same-data, same-protocol pairwise comparison at a scale (82,255 training emails) substantially larger than the closest prior benchmark, combined with LIME, SHAP, and attention-based analysis to help interpret why the two model families fail.

## 2. Related Work

Lexical, frequency-based classifiers such as TF-IDF combined with Logistic Regression remain a strong and interpretable baseline for text classification because Logistic Regression handles the very large, sparse feature spaces that TF-IDF vectorisation produces (Joachims, 1998). Transformer-based transfer learning, exemplified by DistilBERT (Sanh et al., 2019), instead learns contextual token representations from large pretraining corpora before fine-tuning on a target task, and has generally out-performed lexical baselines on clean benchmarks. Adversarial machine learning research characterises attacks against such models along axes including attacker knowledge (white-box vs black-box), attack timing (poisoning vs evasion), and attack goal (Biggio and Roli, 2017); the test-time, black-box evasion setting  an attacker who can reword a message but has no access to model internals  is the most realistic threat model for deployed phishing filters and is the one adopted here. Adversarial NLP work such as Alzantot et al. (2018) and Jin et al. (2020) has shown that transformer models, including BERT variants, can be evaded through word- and character-level perturbations that preserve meaning to a human reader.

The closest prior work is Gholampour and Verma (2023), whose IWSPA workshop paper introduced the adversarial and synthetic phishing evaluation datasets used in this study. Their adversarial set contains 275 examples generated with TextFooler, PWWS, DeepWordBug, and BAE from the TextAttack framework, combining word- and character-level manipulation strategies, and their synthetic set contains 5,000 programmatically generated phishing emails; both were trained on a smaller corpus of roughly 5,700 emails. This paper builds on that work in two respects: it trains both compared models on a unified corpus of 82,255 emails from six independent sources  roughly fourteen times larger  and it evaluates both model families under identical training data, preprocessing, and evaluation protocol so that any difference in adversarial degradation can be attributed to model architecture rather than to differing experimental conditions. Three gaps motivate this design: much prior work treats clean held-out accuracy as sufficient evidence of model quality; controlled, same-protocol comparisons between lexical and transformer phishing detectors under adversarial conditions remain scarce; and the phishing threat landscape is itself shifting as LLM-generated phishing becomes more fluent and stylistically distinct from historical examples.

## 3. Methodology

### 3.1 Threat Model and Experimental Design

The study follows a comparative, quantitative design rather than an architectural one: the aim was not to build a new phishing detector but to test, under matched conditions, whether model type affects the degree of adversarial degradation. The adopted threat model is a test-time, black-box evasion attacker who can alter an email's wording before submission but has no access to model weights, training data, or confidence scores (Biggio and Roli, 2017) a realistic approximation of how phishing authors iterate against deployed filters. Both models were evaluated on the same three conditions: a normal in-distribution test drawn from the training distribution; a synthetic phishing test measuring generalisation to newly generated phishing-style text; and an adversarial test measuring degradation against text deliberately manipulated to evade detection. The primary robustness measure is the **adversarial robustness drop**, defined as normal-test accuracy minus adversarial-test accuracy.

### 3.2 Datasets

Six publicly available datasets were combined to build the training corpus: the CEAS 2008 phishing dataset, the Enron legitimate-email corpus, the Ling-Spam dataset, the Nazario phishing corpus, the Nigerian Fraud (advance-fee) dataset, and the SpamAssassin corpus. Each was standardised to a common {text, label} schema (subject and body concatenated where both existed; label 1 = phishing, 0 = legitimate), with no stemming or stopword removal so that both the lexical and contextual model could use the full token sequence. The resulting corpus contained 82,255 emails: 42,675 phishing (51.9%) and 39,580 legitimate (48.1%), split 80/20 (stratified, random_state = 42) into 65,804 training and 16,451 held-out test examples. Two further datasets, from Gholampour and Verma (2023), were held out entirely from training and used only for evaluation: the IWSPA synthetic phishing set (5,000 emails) and the IWSPA adversarial set (275 emails: 163 phishing, 112 legitimate), available at github.com/ReDASers/IWSPA-2023-Adversarial-Synthetic-Dataset.

### 3.3 Models

The TF-IDF + Logistic Regression model used a vectoriser capped at 20,000 terms over unigrams and bigrams, with Logistic Regression parameters max_iter = 1000, C = 1.0, and solver = “lbfgs”; training completed in under 90 seconds. DistilBERT (Sanh et al., 2019) was fine-tuned from the distilbert-base-uncased checkpoint via the Hugging Face Trainer API on a stratified sample of 50,000 training examples for two epochs with a batch size of 16, using the Apple MPS backend on a single Apple MacBook Pro (M4). A fixed random seed (42) was used throughout for reproducibility.

### 3.4 Evaluation Protocol

Both models were scored on accuracy, precision, recall, F1-score, phishing recall, legitimate recall, and confusion matrices under each of the three test conditions, with the synthetic test excluding legitimate-recall (it contains phishing-only emails). No information from the synthetic or adversarial sets was used during training or model selection. With 275 adversarial examples, one additional misclassification shifts adversarial accuracy by approximately 0.36 percentage points, so small differences between the two models' adversarial scores are interpreted cautiously throughout.

### 3.5 Explainability Methods

Explainability was used to help interpret not merely report the performance results. LIME (Ribeiro, Singh and Guestrin, 2016) was applied to selected adversarial examples to surface word-level contributions to individual predictions. SHAP (Lundberg and Lee, 2017) was used for global feature importance, using LinearExplainer for TF-IDF + LR and PartitionExplainer for DistilBERT. For DistilBERT, attention was additionally examined through [CLS]-token attention, token-to-token heatmaps, and attention rollout across all six transformer layers (Abnar and Zuidema, 2020). These methods are treated as supporting diagnostic evidence rather than proof of causal model reasoning.

## 4. Results

### 4.1 Master Results

Table 1 reports accuracy, precision, F1, class-specific recall, and the adversarial robustness drop for both models across all three test conditions.

***Table 1. Model performance across normal, synthetic, and adversarial test conditions***

| Model | Test Condition | Accuracy | Precision | F1 | Phishing Recall | Legit. Recall | Acc. Drop |
|---|---|---|---|---|---|---|---|
| TF-IDF + LR | Normal | 98.59% | 0.9855 | 0.9864 | 98.73% | 98.43% | |
| TF-IDF + LR | Synthetic | 85.04% | 1.0000 | 0.9192 | 85.04% | N/A | 13.55pp |
| TF-IDF + LR | Adversarial | 64.00% | 0.6455 | 0.7415 | 87.12% | 30.36% | 34.59pp |
| DistilBERT | Normal | 99.04% | 0.9936 | 0.9907 | 98.78% | 99.32% | |
| DistilBERT | Synthetic | 88.98% | 1.0000 | 0.9417 | 88.98% | N/A | 10.06pp |
| DistilBERT | Adversarial | 63.64% | 0.6352 | 0.7475 | 90.80% | 24.11% | 35.40pp |

Both models performed strongly on clean, in-distribution data (98.59% and 99.04% accuracy respectively) and generalised reasonably well to synthetic phishing (85.04% and 88.98%). Under adversarial testing, however, both collapsed to approximately 64% accuracy: TF-IDF + LR dropped 34.59 percentage points, and DistilBERT dropped 35.40 percentage points a difference of only 0.36 percentage points between the two models' adversarial accuracy, equivalent to one email in the 275-sample adversarial set.

## 4.2 Comparative Robustness Analysis

***Table 2. Key robustness comparison, TF-IDF + LR vs DistilBERT***

| Metric | TF-IDF + LR | DistilBERT | Interpretation |
|---|---|---|---|
| Normal accuracy | 98.59% | 99.04% | DistilBERT marginally higher (+0.45pp) |
| Synthetic accuracy | 85.04% | 88.98% | DistilBERT more generalisable (+3.94pp) |
| Adversarial accuracy | 64.00% | 63.64% | Effectively equivalent (0.36pp gap ≈ 1 email) |
| Adversarial robustness drop | 34.59pp | 35.40pp | No meaningful difference (0.81pp gap) |
| Phishing recall (adv.) | 87.12% | 90.80% | DistilBERT detects slightly more phishing |
| Legitimate recall (adv.) | 30.36% | 24.11% | TF-IDF + LR produces fewer false positives |

DistilBERT's advantage on normal and synthetic data did not carry over to the adversarial condition, contrary to the initial expectation grounded in the adversarial-NLP literature that contextual representations would provide a clear robustness advantage. The two models also failed differently: DistilBERT retained slightly higher phishing recall (90.80% vs 87.12%) but produced more false positives, with legitimate recall falling to 24.11% against 30.36% for TF-IDF + LR.

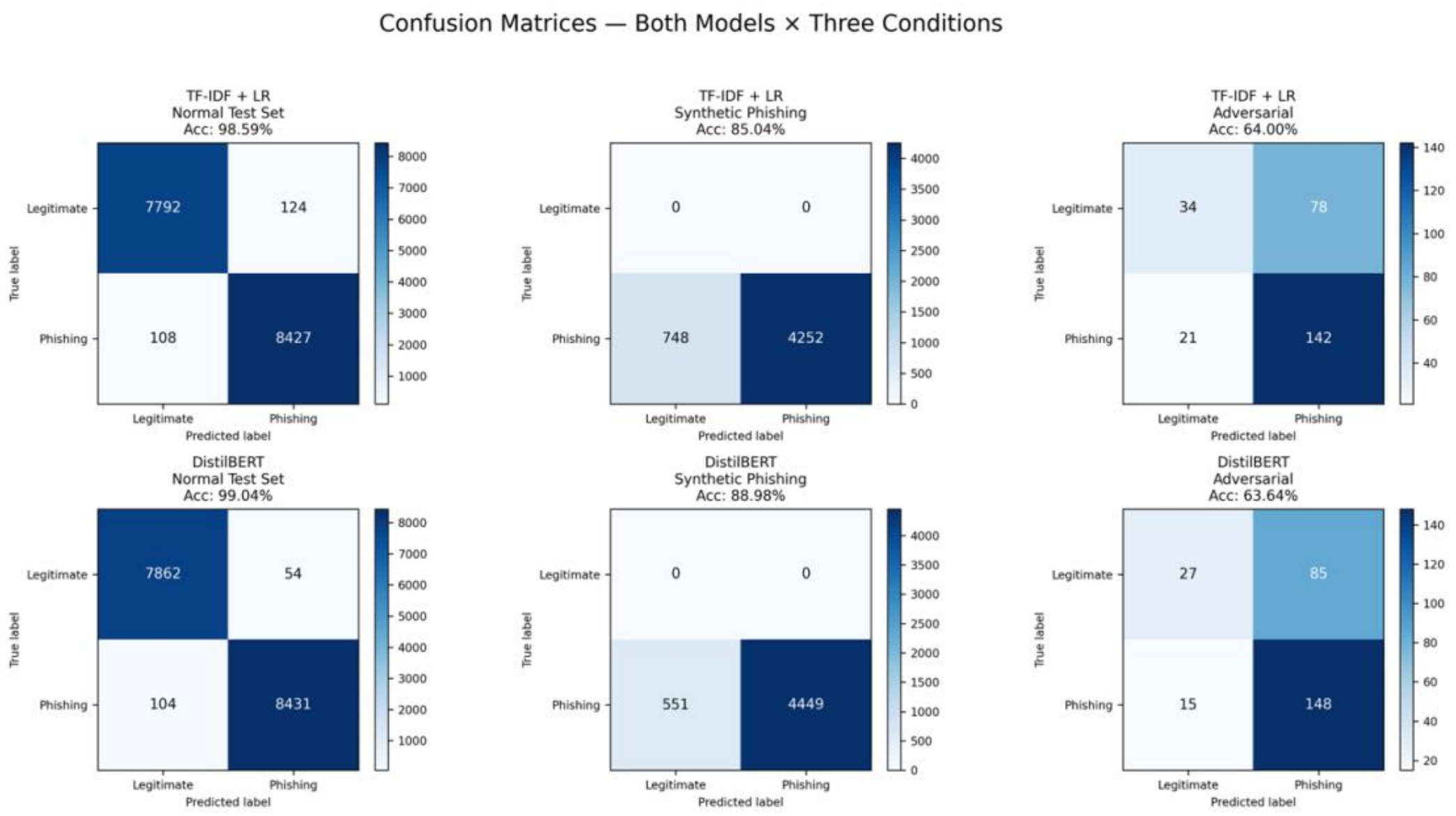

***Figure 1. Confusion matrices for both models across all three evaluation conditions.***

## 4.3 Pairwise Error Analysis

Because both models were evaluated on the identical 275-email adversarial set, sample-level agreement could be examined directly. Table 3 shows that the two models agreed on the large majority of samples 54.9% correct by both, 27.3% wrong for both  but each also made a comparable number of exclusive errors (24 for TF-IDF + LR, 25 for DistilBERT), indicating that although overall adversarial accuracy was almost identical, the models did not fail on exactly the same examples.

***Table 3. Pairwise error agreement between models on the adversarial set***

| Category (adversarial set, n = 275) | Count | Share | Interpretation |
|---|---|---|---|
| Both models correct | 151 | 54.9% | Reliable core both models handled |
| Both models wrong | 75 | 27.3% | Hard adversarial cases neither model handled |
| Only TF-IDF + LR wrong | 24 | 8.7% | Cases where DistilBERT's context helped |
| Only DistilBERT wrong | 25 | 9.1% | Cases where TF-IDF's lexical cues helped |

The error profile is also asymmetric by class. TF-IDF + LR misclassified 78 of 112 legitimate emails as phishing (a 69.64% adversarial false-positive rate) and missed 21 of 163 phishing emails (12.88% false-negative rate); DistilBERT misclassified 85 of 112 legitimate emails (75.89% false-positive rate) and missed 15 phishing emails (9.20% false-negative rate). Both models therefore retained relatively high phishing recall under attack but suffered a much larger collapse in legitimate recall the operational risk in this setting is not only missed phishing but a potential flood of false positives.

## 5. Discussion

The central finding is that clean-data accuracy did not predict adversarial robustness: both models exceeded 98% accuracy normally yet converged to almost identical performance (≈64%) once inputs were adversarially manipulated, and the 0.81-percentage-point gap between their robustness drops is well within the noise expected from a 275-sample test set. This does not mean the two models behaved identically  LIME and SHAP analysis of the adversarial examples showed no overlap in top-ranked features between models: TF-IDF + LR's explanations were dominated by visible lexical and structural cues (e.g. “click”, “account”, “http”, “thanks”), consistent with a bag-of-words representation, while DistilBERT's explanations were more dispersed across context-dependent tokens and named entities. Attention-rollout analysis similarly showed DistilBERT's decisions were not concentrated on obvious phishing keywords but distributed across special tokens ([CLS], [SEP]) and ordinary words. Different evidence, in other words, produced a similar failure mode which is itself informative: it suggests the adversarial perturbations in this dataset degrade both lexical and contextual signal simultaneously, rather than exploiting a weakness specific to one representation.

These results have a direct cybersecurity risk-management implication. For an illustrative organisation processing 50,000 emails per day at a 2% phishing rate (≈ 1,000 phishing, 49,000 legitimate emails), the adversarial legitimate-recall figures observed here (30.36% for TF-IDF + LR, 24.11% for DistilBERT) would translate into a very large volume of false-positive alerts relative to true phishing volume  a burden that could overwhelm SOC triage capacity even though phishing recall itself remained comparatively high. This example is illustrative rather than a deployment forecast, since the adversarial test set is small and does not reflect real-world class balance, but it underlines why procurement and deployment decisions for email-security tooling should weigh adversarial robustness and false-positive impact alongside clean-benchmark accuracy.

## 6. Limitations

- The adversarial test set contains only 275 samples, limiting the statistical power of fine-grained model-to-model comparisons; one misclassification shifts adversarial accuracy by ~0.36 percentage points.
- DistilBERT was fine-tuned on a 50,000-example capped sample of the 65,804 available training examples because of hardware constraints, rather than the full training set.
- Results are based on a single random seed and were not averaged across multiple runs.
- SHAP's PartitionExplainer for DistilBERT is computationally expensive, limiting the number of adversarial emails analysed for that component.
- The training corpus is predominantly English language, which limits how far the findings generalise to non-English phishing.

## 7. Future Work

- Ensemble approaches: the partly non-overlapping exclusive errors (Table 3) suggest combining lexical and transformer-based representations may reduce false positives and false negatives under adversarial conditions; this project does not test that hypothesis.
- Larger adversarial test sets, to improve the statistical reliability of model-to-model robustness comparisons.
- Dynamic attention mechanisms, e.g. following Shen et al. (2024), targeting the distributional vulnerability suggested by the attention-rollout analysis.
- Extending SHAP analysis to quantify feature-interaction effects under adversarial perturbation.

## 8. Conclusion

This study compared a TF-IDF + Logistic Regression model and a fine-tuned DistilBERT model under identical training data and evaluation protocol, across normal, synthetic, and adversarial phishing-email conditions. Both models exceeded 98% accuracy on clean data but converged to approximately 64% accuracy under adversarial testing a 34.59-percentage-point drop for TF-IDF + LR and a 35.40-percentage-point drop for DistilBERT, a difference too small (0.36pp, one email) to support a claim that either model is meaningfully more robust. The main conclusion is therefore not that one model outperforms the other, but that clean-data accuracy did not predict adversarial robustness in this setting, despite the intuitive, literature-grounded expectation that DistilBERT's contextual representations would provide a clear advantage. Explainability analysis indicates the two models relied on different evidence while still exhibiting similar vulnerability, and pairwise error analysis shows their failures were only partly overlapping both findings that motivate future ensemble work. For cybersecurity practice, the results support treating adversarial testing, explainability analysis, and operational false-positive impact as standard components of phishing-detection evaluation, rather than relying on clean benchmark accuracy alone.

## Ethics Statement

All datasets used are publicly available research corpora; no new human-participant data was collected, and no individuals were identified from email content. All processing was performed in an isolated local environment and was not shared or transferred. The models developed are detection-oriented classifiers only and provide no generative capability that could be used to produce phishing content. Formal human-participant ethics approval was not required under the authors' institutional ethics policy on this basis.

## Data Availability Statement

The six training-corpus sources (CEAS 2008, Enron, Ling-Spam, Nazario, Nigerian Fraud, SpamAssassin) and the IWSPA synthetic and adversarial evaluation sets (Gholampour and Verma, 2023) are publicly available, including at github.com/ReDASers/IWSPA-2023-Adversarial-Synthetic-Dataset and kaggle.com/datasets/naserabdullahalam/phishing-email-dataset. Trained model artefacts and analysis notebooks are available from the corresponding authors on reasonable request.

## Author Contributions

Tanveer Ahmed: conceptualisation, data curation, methodology, software, formal analysis, investigation, visualisation, writing. Seyedali Pourmoafi: supervision, conceptualisation, methodology review, writing review and editing.